\documentclass[letterpaper, 10 pt, conference]{ieeeconf} 
\usepackage{cite}
\usepackage{amsmath,amssymb,amsfonts}
\usepackage{algorithmic}
\usepackage{graphicx}
\usepackage{xurl}

\IEEEoverridecommandlockouts                              

\title{\LARGE \bf
Hugging Suit: \\Pneumatically-Actuated System Design for Remote Haptic Experiences}

\author{Russian (Ruo-Xuan) Wu$^{1}$, Luke Hespanhol$^{2}$, Marius Hoggenmueller$^{2}$, \\Hannes Waldschütz$^{3}$, and Eva Hornecker$^{3}$
\thanks{$^{1}$Formerly with Bauhaus-Universität Weimar, 99423 Germany. Now with the Design Lab, The University of Sydney, NSW 2008, Australia. Emails:\texttt{russian.wu@sydney.edu.au}}%
\thanks{$^{2}$Design Lab, The University of Sydney, NSW 2008, Australia. Emails: }%
\thanks{\texttt{luke.hespanhol, marius.hoggenmueller@sydney.edu.au}}%
\thanks{$^{3}$Bauhaus-Universität Weimar, 99423 Germany. Emails:}%
\thanks{\texttt{hannes.waldschuetz, eva.hornecker@uni-weimar.de}}%
}

\begin{document}

\maketitle
\thispagestyle{empty}
\pagestyle{empty}

\section{Introduction}
    The COVID-19 pandemic emphasized the importance of remote emotional communication and highlighted a gap in existing technologies that lack haptic channels. While video calls maintain visual and auditory presence, they cannot convey the emotional depth of physical contact, especially hugging, a universally recognized form of intimacy and support~\cite{Barnett_KeepInTouch_2005, Field_ImportanceOfTouch_2004}. To explore this challenge, we developed the \textit{Hugging Suit}, a pneumatically-actuated wearable system that enables users to simulate and receive remote hugs in real time. Unlike prior haptic systems that focus solely on tactile sensation, our approach integrates both technical and experiential considerations~\cite{Mueller_HugOveraDistance_2005}. The system integrates a programmable Air Actuator Matrix (AAM), a portable high-pressure pneumatic unit, and a fabric-based pressure sensor layer, enabling precise, wearable haptic feedback while remaining lightweight and mobile. Guided by a Research through Design (RtD) methodology, we iteratively refined the prototype to improve tactile resolution, overall usability, and user comfort, while introducing modular components to support flexible and scalable haptic configurations~\cite{Zimmerman_Research_2007}. In parallel, we explored how specific contextual factors, such as lighting, privacy, and the visibility of the remote partner, might shape the emotional and perceptual experience of receiving a remote hug. Preliminary feedback suggests that low-lit, private spaces, especially when users could also see their remote partner via video, improved emotional engagement and comfort during mediated haptic experiences. Our low-cost, maker-space-friendly approach contributes to affective haptics applications in long-distance relationships, emotional regulation, and remote therapeutic support.

\begin{figure*}[htbp]
  \centering
  \includegraphics[width=\linewidth]{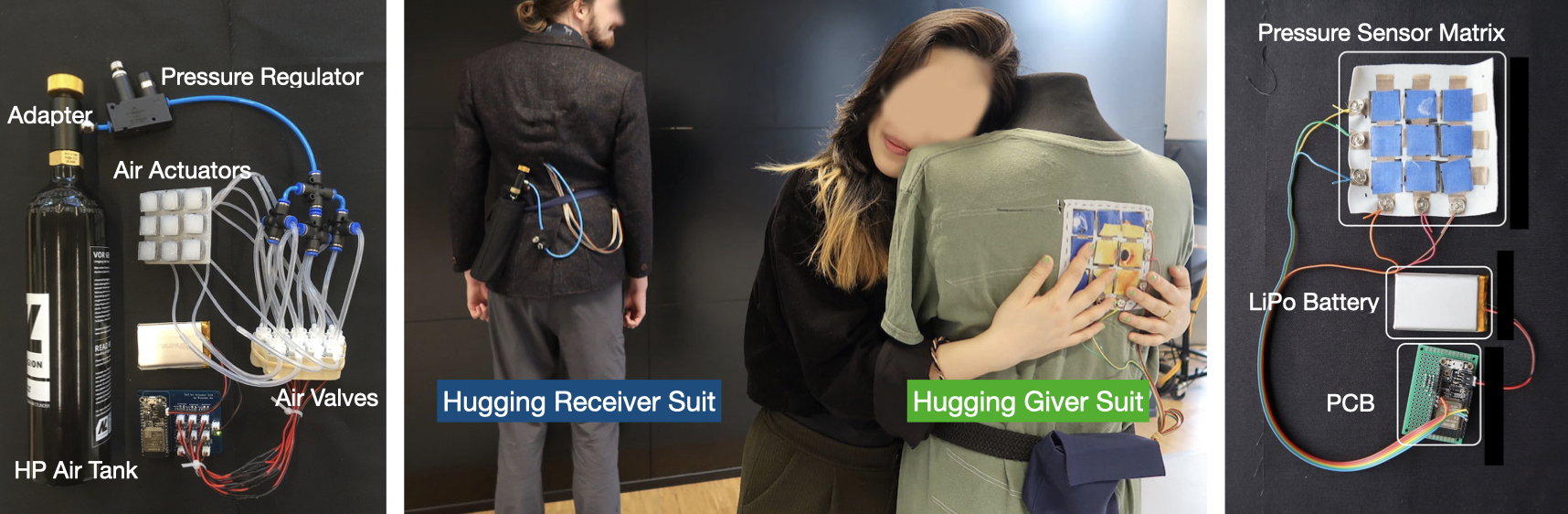}
  \caption{\textbf{Hugging Suit: System Overview and Modular Components.} Diagram of the Hugging Receiver and Giver Suits, highlighting core components: AAM, PSM, and the pneumatic system enabling remote hugging interaction.}
  \label{fig:Overview}
\end{figure*}

\section{System Design}

    The \textit{Hugging Suit} is a pneumatically-actuated wearable system designed to simulate remote hugs through programmable, body-conforming tactile actuation. It comprises three main components: an Air Actuator Matrix (AAM), a Pressure Sensor Matrix (PSM), and a high-pressure pneumatic unit adapted from a refillable paintball air tank, which enables mobile deployment of the suit without bulky compressors. These components are integrated into a modular system supporting comfort and maker-space fabrication. The current prototype implements a single back-mounted module, with actuators positioned to reflect typical hug contact points in that region. The system’s modular architecture is designed to support future extension to additional body areas such as the shoulders and chest.

    The system consists of two garments: the Hugging Giver Suit, which incorporates the Pressure Sensor Matrix (PSM) to detect the act of hugging; and the Hugging Receiver Suit, which integrates the Air Actuator Matrix (AAM) and pneumatic components to reproduce the tactile sensation (see Fig.~\ref{fig:Overview} on the next page). Communication between the two suits is managed via a wireless OSC protocol with minimal communication latency, enabling responsive remote interaction.

    The AAM, built using McKibben artificial muscles, delivers tactile feedback through a grid of soft pneumatic actuators installed along the back. The spacing of actuators was informed by known two-point discrimination values to optimize coverage while minimizing weight~\cite{Nolan_Quantitative_1985}. Each actuator was cast in silicone using custom 3D-printed molds, following a multi-stage fabrication process adapted from prior soft robotics work~\cite{Gohlke_Pneumatibles_2016}.
    
    The PSM is constructed as a textile-based piezoresistive sensor array using stacked layers of conductive and resistive fabrics. This configuration allows the system to detect pressure changes associated with hug initiation and contact intensity. Designed for flexibility and wearer comfort, the sensors are embedded into the garment structure using a laminated layering technique. Connection points are sewn with fasteners to enable reliable integration with the actuator control system. This sensing layer enables real-time response while maintaining the softness and conformity expected of a wearable garment.
    
    A notable feature of the \textit{Hugging Suit} is its use of a high-pressure pneumatic system powered by a refillable paintball air tank. This configuration helps reduce the noise, weight, and bulk compared to conventional air pumps, making the system more suitable for untethered use~\cite{Delazio_ForceJacket_2018}. In the current setup, the air tank is filled to 8 bar and can operate the actuator matrix continuously for up to 5 minutes per charge. Actuators operate in binary (on/off) states, with solenoid valves controlled by a microcontroller-based system. This setup enables simple, fast-response actuation and can be replicated using accessible fabrication techniques, supporting ongoing prototyping and iterative refinement in maker-space settings. This prototype also served as the basis for a user study investigating how contextual factors shape remote hugging experiences.

\section{Preliminary Results}
    In this early phase, we gathered experiential feedback through two exploratory studies. Participants interacted with functional prototypes in staged environments. We aimed to evaluate the \textit{Hugging Suit}’s tactile expressiveness and contextual responsiveness, gathering early feedback on comfort, perceived realism, and emotional impact.
    
    Participants generally responded positively to the concept and tactile actuation. Over 90\% reported that they could clearly feel the haptic feedback. However, less than half (44.1\%) agreed that the sensation resembled a real hug. Reported gaps were attributed to limited coverage, lack of warmth, and insufficient softness. While realism remains a challenge, participants appreciated the comfort and wearability of the system, suggesting that the modular garment-based design supports future improvement.
    
    Contextual factors also shaped user experiences. Participants preferred low-lit, private settings and reported increased emotional resonance when seated comfortably or when the hug giver was visible via video call. Several participants described lingering tactile sensations after removing the suit, which they associated with comfort and emotional carryover. Others proposed possible future uses, including stress relief, connecting with distant family members, or enhancing video calls with a sense of physical presence. These insights inform next steps in refining the system's realism, responsiveness, and contextual fit.

\section{Conclusion and Next Steps}

    This work presents the \textit{Hugging Suit}, a pneumatically-actuated wearable system that simulates remote hugs through modular, body-conforming tactile actuation. Developed through a Research through Design (RtD) process, the system integrates an Air Actuator Matrix (AAM), a textile-based Pressure Sensor Matrix (PSM), and a high-pressure pneumatic unit adapted from a refillable air tank, enabling embodied haptic communication in remote contexts.

    Preliminary findings highlight the importance of context-aware design in shaping mediated touch experiences. Participants reported stronger emotional engagement in low-lit, private environments, particularly when the hug giver was visible. They also appreciated the system’s overall comfort and emotional potential while identifying opportunities for improvement, including tactile realism, actuator coverage, and wearability for extended use. We believe the \textit{Hugging Suit} represents a valuable step toward affective haptic systems that support emotional regulation, remote intimacy, and context-sensitive interaction in everyday environments.
    



\bibliographystyle{IEEEtran}
\bibliography{sample-base}

@inproceedings{Delazio_ForceJacket_2018,
author = {Delazio, Alexandra and Nakagaki, Ken and Klatzky, Roberta L. and Hudson, Scott E. and Lehman, Jill Fain and Sample, Alanson P.},
title = {Force Jacket: Pneumatically-Actuated Jacket for Embodied Haptic Experiences},
year = {2018},
isbn = {9781450356206},
publisher = {Association for Computing Machinery},
address = {New York, NY, USA},
url = {https://doi.org/10.1145/3173574.3173894},
doi = {10.1145/3173574.3173894},
booktitle = {Proceedings of the 2018 CHI Conference on Human Factors in Computing Systems},
pages = {1–12},
numpages = {12},
location = {<conf-loc>, <city>Montreal QC</city>, <country>Canada</country>, </conf-loc>},
series = {CHI '18}
}

@article{Barnett_KeepInTouch_2005,
  title={Keep in touch: The importance of touch in infant development},
  author={Barnett, Lynn},
  journal={Infant Observation},
  volume={8},
  number={2},
  pages={115-123},
  year={2005},
  publisher={Taylor \& Francis}
}

@book{Field_ImportanceOfTouch_2004,
	title = {The Importance of Touch},
	volume = {67},
	url = {https://misc.karger.com/gazette/67/Field/art_4.htm},
	pagetotal = {10-12},
	publisher = {Karger Gazette},
	author = {Field, Tiffany},
	urldate = {2022-05-03},
	year = {2004},
    address = {Miami, Florida.},
}

@inproceedings{Mueller_HugOveraDistance_2005,
author = {Mueller, Florian 'Floyd' and Vetere, Frank and Gibbs, Martin R. and Kjeldskov, Jesper and Pedell, Sonja and Howard, Steve},
title = {Hug over a Distance},
year = {2005},
isbn = {1595930027},
publisher = {Association for Computing Machinery},
address = {New York, NY, USA},
url = {https://doi.org/10.1145/1056808.1056994},
doi = {10.1145/1056808.1056994},
booktitle = {CHI '05 Extended Abstracts on Human Factors in Computing Systems},
pages = {1673–1676},
numpages = {4},
location = {Portland, OR, USA},
series = {CHI EA '05}
}

@article{Nolan_Quantitative_1985,
        title = {Quantitative Measure of Cutaneous Sensation},
        volume = {65},
	issn = {0031-9023, 1538-6724},
	url = {https://academic.oup.com/ptj/article-lookup/doi/10.1093/ptj/65.2.181},
	doi = {10.1093/ptj/65.2.181},
	pages = {181--185},
	number = {2},
	journaltitle = {Physical Therapy},
	author = {Nolan, Michael F.},
	urldate = {2022-03-24},
	year = {1985},
    month = feb,
	langid = {english},
}

@inproceedings{Zimmerman_Research_2007,
author = {Zimmerman, John and Forlizzi, Jodi and Evenson, Shelley},
title = {Research through Design as a Method for Interaction Design Research in HCI},
year = {2007},
isbn = {9781595935939},
publisher = {Association for Computing Machinery},
address = {New York, NY, USA},
url = {https://doi.org/10.1145/1240624.1240704},
doi = {10.1145/1240624.1240704},
booktitle = {Proceedings of the SIGCHI Conference on Human Factors in Computing Systems},
pages = {493–502},
location = {<conf-loc>, <city>San Jose</city>, <state>California</state>, <country>USA</country>, </conf-loc>},
series = {CHI '07}
}

@inproceedings{Gohlke_Pneumatibles_2016,
author = {Gohlke, Kristian and Hornecker, Eva and Sattler, Wolfgang},
title = {Pneumatibles: Exploring Soft Robotic Actuators for the Design of User Interfaces with Pneumotactile Feedback},
year = {2016},
isbn = {9781450335829},
publisher = {Association for Computing Machinery},
address = {New York, NY, USA},
url = {https://doi.org/10.1145/2839462.2839489},
doi = {10.1145/2839462.2839489},
booktitle = {Proceedings of the TEI '16: Tenth International Conference on Tangible, Embedded, and Embodied Interaction},
pages = {308–315},
numpages = {8},
location = {Eindhoven, Netherlands},
series = {TEI '16}
}

\end{document}